\documentclass[aps,prl,twocolumn,superscriptaddress]{revtex4-2}

\usepackage{graphicx}
\usepackage{dcolumn}
\usepackage{bm}
\usepackage[usenames,dvipsnames]{color} 
\usepackage{physics}
\usepackage{siunitx}
\usepackage[colorlinks,urlcolor=blue,citecolor=blue,linkcolor=blue]{hyperref}
\usepackage{ulem} 
\usepackage{nicefrac}
\usepackage[all]{nowidow}
\usepackage{comment}

\begin{document}

\preprint{APS/123-QED}
\title{Realization of repulsive polarons in the strongly correlated regime}

\author{R. Henke}
\affiliation{Institute for Quantum Physics, Universität Hamburg, Luruper Chaussee 149, 22761 Hamburg, Germany}

\author{J. Levinsen}
\affiliation{School of Physics and Astronomy, Monash University, Victoria 3800, Australia}

\author{M. M. Parish}
\affiliation{School of Physics and Astronomy, Monash University, Victoria 3800, Australia}

\author{J. Boronat}
\affiliation{Departament de Física, Campus Nord B4-B5, Universitat Politècnica de Catalunya, E-08034 Barcelona, Spain}

\author{G. E. Astrakharchik}
\affiliation{Departament de Física, Campus Nord B4-B5, Universitat Politècnica de Catalunya, E-08034 Barcelona, Spain}

\author{H. Moritz}
\email{hennning.moritz@uni-hamburg.de}
\affiliation{Institute for Quantum Physics, Universität Hamburg, Luruper Chaussee 149, 22761 Hamburg, Germany}
\affiliation{The Hamburg Centre for Ultrafast Imaging, Universität Hamburg, Luruper Chaussee 149, 22761 Hamburg, Germany}

\author{C. R. Cabrera}
\email{cesar.cabrera@uni-hamburg.de}
\affiliation{Institute for Quantum Physics, Universität Hamburg, Luruper Chaussee 149, 22761 Hamburg, Germany}
\affiliation{The Hamburg Centre for Ultrafast Imaging, Universität Hamburg, Luruper Chaussee 149, 22761 Hamburg, Germany}

\date{\today}

\begin{abstract}
Mobile impurities interacting with a quantum medium form quasiparticles known as polarons, a central concept in many-body physics. While the quantum impurity problem has been extensively studied with ultracold atomic gases, repulsive polarons in the strongly correlated regime have remained elusive. Typically, the impurity atoms bind into molecules or rapidly decay into deeper lying states before they can acquire an appreciable dressing cloud. Here, we report on the realization of polarons in a strongly repulsive quasi-two-dimensional quantum gas. Using a superfluid of $^6$Li dimers, we introduce impurities by promoting a small fraction of the dimers into higher levels of the transverse confining potential. These novel synthetic-spin polarons give access to the strongly repulsive regime where common decay channels are suppressed. We extract key polaron properties---the energy, quasiparticle residue, and effective mass---using trap modulation and Bragg spectroscopy. Our measurements are well captured by a microscopic T-matrix approach and quantum Monte Carlo simulations, revealing deviations from mean-field predictions. In particular, we measure a significant enhancement of the polaron mass, with values exceeding twice the free dimer mass. Our demonstration of a stable repulsive Bose polaron establishes a platform for studying impurity physics in low-dimensional and strongly correlated systems.
\end{abstract}

\maketitle

Polarons---quasiparticles formed by impurities interacting with a surrounding quantum medium---play a crucial role in understanding quantum many-body systems.
Originally introduced by Landau and Pekar to describe electrons coupled to a crystal lattice~\cite{LandauPekar}, the polaron concept 
has since provided valuable insights into phenomena ranging from the phase diagram of superfluid helium mixtures~\cite{Bardeen1967} to the optical response of doped semiconductors~\cite{Sidler_NatPhys_2017},
and the pairing mechanism underlying high-temperature superconductivity~\cite{Lee2006}. 
Despite its conceptual simplicity, the quantum impurity problem represents a paradigmatic many-body challenge, and a central question is how the quasiparticle picture evolves in the strongly correlated regime. 

Ultracold atoms provide an ideal platform for studying polaron physics due to the ability to tune the atom-atom interactions~\cite{Chin2010-jc} and prepare impurities in bosonic and fermionic mediums~\cite{Massignan_RPP2014,Scazza2022}. For both these Bose and Fermi polarons, it is now well established that the impurity spectrum features attractive and repulsive branches, where the impurities attract or repel the surrounding medium~\cite{Massignan_RPP2014,Scazza2022}.  However, this scenario involves underlying attractive interactions where the physics of the attractive polaron is influenced by few-atom bound states~\cite{Schirotzek2009,Koschorreck2012,Hu2016,Jorgensen2016,Oppong2019,Yan2019,etrych2024}, while the repulsive polaron is intrinsically unstable~\cite{Kohstall2012,Koschorreck2012,Hu2016,Jorgensen2016,Scazza2017,Oppong2019,etrych2024,Morgen2025}. The situation is particularly severe for Bose polarons in the strongly interacting regime, since bosons tend to form clusters and are more susceptible to three-body losses~\cite{Levinsen2015,Naidon2017}. As a result, experimental studies of repulsive polarons have so far been limited to transient regimes and it remains unclear 
whether the quasiparticle picture holds under strong repulsion~\cite{Scazza2022}.

\begin{figure}[tbh!]
\includegraphics[scale=1]{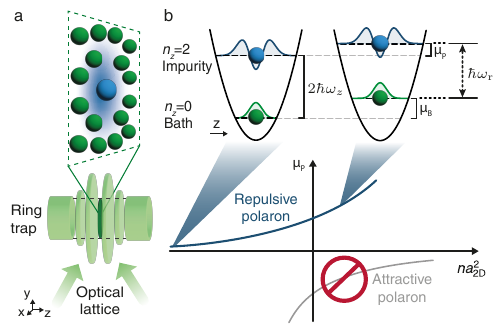}
\caption{\textbf{Experimental realization of repulsive synthetic-spin polarons.} a) An impurity (blue) is immersed in a homogeneous 2D Bose-Einstein condensate (green). The atomic cloud is confined to a single 2D plane of a blue-detuned optical lattice in combination with a ring trap (box potential in the radial plane).
b) The eigenstates of the harmonic trapping potential along the $z$ direction play the role of synthetic spins, 
allowing us to realize a 2D repulsive polaron.  
Here, the bath corresponds to a superfluid of density $n$ in the ground state $n_z=0$, while the impurity is created by populating a higher level such as $n_z=2$. 
Without interactions, the energy $\hbar\omega_\textrm{r}$ to address this transition is $2\hbar\omega_z$, the harmonic oscillator level spacing. With increasing repulsion $na_\textrm{2D}^2$, $\hbar\omega_\textrm{r}$ exhibits an interaction shift arising from the difference between the bath chemical potential $\mu_\mathrm{B}$ and the repulsive polaron energy $\mu_\mathrm{P}$. 
In contrast to previous Bose polaron realizations, there is no underlying attractive branch.}
\label{Fig:figure1}
\end{figure}

Here, we experimentally realize a stable repulsive Bose polaron in the strongly interacting regime, overcoming the intrinsic metastability observed in previous studies~\cite{Hu2016,Jorgensen2016,etrych2024,Morgen2025}. 
We focus on the two-dimensional (2D) geometry (Fig.~\ref{Fig:figure1}a), where correlations are enhanced and striking phenomena are predicted, such as effective mass divergence and the breakdown of the quasiparticle picture~\cite{Grusdt2016,Pastukhov_2018,Ardila2020,Tajima2021,CardenasCastillo2023,Nakano2024}.
Our approach employs bosonic Feshbach molecules consisting of pairs of fermions~\cite{Zwerger2011} which have a strong immunity against typical three-body decay due to the Pauli exclusion principle~\cite{Petrov2004}.
In addition, they feature repulsive interactions~\cite{Petrov2004} with large tunability, allowing us to realize strongly repulsive superfluids. 
Simultaneously, the strong 2D confinement enables us to create impurities within the superfluid by exciting a small number of dimers from the ground state to an excited state of the trapping potential, with the quantum number $n_z$ acting as a synthetic spin (Fig.~\ref{Fig:figure1}b).
This mapping of impurity and bath into different vibrational states gives rise to a new class of quasiparticle which we term the \textit{synthetic-spin polaron}, characterized by purely repulsive interactions and the absence of any underlying attractive branch. 
We access the polaron spectral function and extract the polaron energy using trap modulation spectroscopy.
Most notably, we measure the effective mass of the Bose polaron for the first time, observing values exceeding twice the bare mass of a non-interacting impurity, a hallmark of strong many-body dressing.
Comparisons with quantum Monte Carlo (QMC) and T-matrix theories reveal pronounced beyond-mean-field effects in a regime previously inaccessible.

The experiments are performed in a homogeneous gas of bosonic dimers, each composed of two $^6$Li fermions occupying the lowest two hyperfine states.
To ensure the bosonic character of the system, we stay in the regime where the dimer size is smaller than the interparticle distance~\cite{supmat}.
The cloud is confined in a highly anisotropic geometry with a trapping frequency $\omega_z = 2 \pi \cdot 11.4\,\mathrm{kHz}$ along the vertical direction, and a uniform box potential providing horizontal confinement~\cite{supmat,Hueck2018-cs, Sobirey2021-xy}, as illustrated in Fig.~\ref{Fig:figure1}a.
The typical temperatures measured after an adiabatic ramp to low interactions~\cite{supmat} are $\widetilde{T}\leq0.02\, T_\mathrm{F}$, where $T_F$ is the Fermi temperature.
We tune the interaction strength via a broad Feshbach resonance \cite{Hueck2018-cs, Sobirey2021-xy}. It is quantified using the dimensionless gas parameter $na_\mathrm{2D}^2$, where $n$ is the total 2D dimer density and the scattering length $a_\mathrm{2D}$ is obtained by mapping the 3D interaction between composite dimers~\cite{Petrov2004} to the quasi-2D geometry~\cite{Petrov2001-kn,supmat}.
The gas parameter also defines the dimensionless parameter $\ln(k_n a_\mathrm{2D})$, which is commonly used to describe 2D systems, where $k_n=\sqrt{4\pi n}$ is the characteristic momentum scale.
We note that in our scheme, the underlying microscopic interaction potential between the impurity and bath particles is the same as that between particles in the bath, with a difference in the effective interactions due to a reduced transverse overlap~\cite{supmat}.
This effect is conceptually similar to the scenario of $^3$He impurities in superfluid $^4$He~\cite{Bardeen1967}, where the larger volume occupied by the lighter $^3$He atom yields different effective interactions despite the same underlying interaction potential.

To probe the properties of a 2D Bose polaron, we measure its spectral response using trap modulation spectroscopy~\cite{Cabrera2025}, which we show~\cite{supmat} is equivalent to injection radio-frequency (RF) spectroscopy widely used in Bose polaron experiments~\cite{Jorgensen2016,etrych2024}.
Specifically, we modulate the vertical trapping frequency as $\omega_z(t) = \omega_z (1 + \alpha \sin{(\omega_{\textrm{m}}}t))$, with modulation frequency $\omega_{\textrm{m}}$ and a small amplitude of $\alpha \leq 0.5\,\%$. This transfers a small fraction of dimers from the superfluid ($n_z=0$) into the $n_z = 2$ harmonic oscillator level, acting as impurities populating the repulsive polaron branch.
Following the excitation, a $\SI{20}{ms}$ hold time allows for collisions between impurities and the bath, resulting in energy redistribution and measurable heating~\cite{supmat}.
The heating rate is directly linked to the zero-momentum spectral function $A_2(\omega)$ for the $n_z=2$ impurities~\cite{supmat}:
\begin{equation}\label{eq:dEdt}
    \frac{dE}{dt} \propto  A_2(\omega_\mathrm{m} + \mu_\mathrm{B}) \, .
\end{equation}
This mapping demonstrates that trap modulation provides the spectral information of injected $n_z=2$ impurities (polarons) into the condensate, and consequently that eigenstates of the trap can be exploited as synthetic spin~\cite{Shashi2014, Weizhe_PRA2020}. 
Note that this differs from previous experiments on trapped motional states in a quantum bath, which only explored weak interactions and required a different species of atom for the impurity~\cite{Scelle2013, rentrop2016}.

Experimentally, we quantify the heating rate via the response function $R(\omega_{\textrm{m}}) = 1-C(\omega_\textrm{m})/C(0)$.
Here, $C(\omega_{\textrm{m}})$ corresponds to the amplitude of low-momentum atoms, determined from the bimodal momentum distribution after time-of-flight imaging.
It is normalized by the unperturbed amplitude $C(0)$, such that $R(\omega_{\textrm{m}})$ becomes non-zero when the modulation is resonant with an excitation.
In the limit of small energy transfers, $R(\omega_{\textrm{m}})$ is proportional to $dE/dt$, and according to Eq.~\eqref{eq:dEdt}, proportional to the polaron spectral function.

\begin{figure}[htb!]
\includegraphics[width=\linewidth]{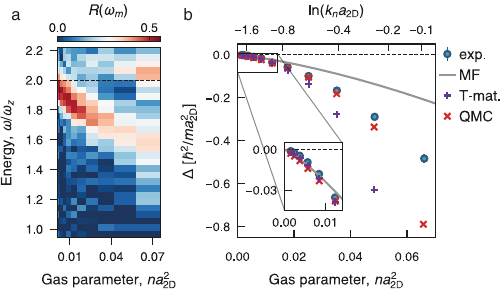}
\caption{\textbf{Polaron energies from trap modulation spectroscopy.} a) Response $R(\omega_{\textrm{m}})$ measured via trap modulation spectroscopy. A well-defined mode is visible, corresponding to the creation of polarons in the $n_z=2$ state.
b) Energy shift $\Delta$, extracted from Lorentzian fits to the spectra shown in panel (a). The error bars of the experimental data are small compared to the symbol size. A comparison with QMC theory shows excellent agreement up to $na_\textrm{2D}^2\lesssim 0.05$ in the strongly interacting regime. Both T-matrix and QMC theories deviate for large interactions, where the details of the scattering potential become relevant, while MF theory fails already for moderate interactions. For small interactions, all theories agree well with the experiment (inset).
}
\label{Fig:figure2}
\end{figure}

In Fig.~\ref{Fig:figure2}a, we show $R(\omega_{\textrm{m}})$ as a function of the gas parameter.
The excitation spectrum features a well-defined resonance across the entire interaction range, clearly below the bare transition $2\omega_z$ (dashed line).  
The interaction shift $\Delta=\hbar\left(\omega_\mathrm{r}-2\omega_z\right)$, shown in Fig.~\ref{Fig:figure2}b, consists of two contributions: the chemical potential of the bath $\mu_\mathrm{B}$, representing the energy lost by removing a particle from the condensate, and the polaron energy $\mu_\mathrm{P}$, corresponding to the energy required to add an impurity into the system. 
Their difference yields the total energy shift: $\Delta=\mu_\mathrm{P}-\mu_\mathrm{B}$. 
Notably, due to the reduced spatial overlap between the impurity and the bath, we find that $\mu_\mathrm{B}>\mu_\mathrm{P}$, and hence $\Delta<0$, even though the polaron is repulsive, $\mu_\mathrm{P}>0$. Since $\mu_\mathrm{B}$ has been previously calculated in weakly interacting 2D Bose gases~\cite{Schick71,Astrakharchik2009,Hadzibabic2011}, our measurement of $\Delta$ provides experimental access to the polaron energy $\mu_\mathrm{P}$, revealing the many-body dressing of the impurity~\cite{supmat}.

To test the validity of the polaron description, we compare our experimental results with two complementary theoretical approaches---diffusion Monte Carlo simulations (denoted as QMC) and a $T$-matrix-based polaron theory~\cite{supmat}---which highlight different aspects of the impurity-bath interactions. Specifically, the $T$-matrix theory includes the possibility of excited impurities undergoing exchange with the bath particles while neglecting the finite range of the interaction potential (set by the dimer size) as well as multiple excitations of the medium. Conversely, the QMC accounts for the range but treats the $n_z=2$ impurities as distinct from the bath particles, which corresponds to the standard polaron scenario. Indeed, the latter approach is reasonable since exchange processes with large energy differences are suppressed for finite-range interactions (further details in~\cite{supmat}). 

The remarkable agreement seen in Fig.~\ref{Fig:figure2}b for $na_\textrm{2D}^2 \lesssim 0.02$ between both theories and the experiment indicates that the polaron picture, indeed, applies to excitations of the transverse degrees of freedom and that the resulting description is universal, as it does not depend on the precise details of the interaction potential.
At stronger interactions, discrepancies between theories and experiment emerge, which can be attributed to several factors. First, the QMC description models the dimers as hard spheres, an approximation that becomes inaccurate when the dimer diameter $a_\mathrm{3D}$ becomes comparable to the oscillator length $l_z$. Additionally, the gas enters the strongly correlated regime when $na_\mathrm{2D}^2 \sim 0.1$, which, for reference, is near the crystallization point predicted for hard disks at $na_\mathrm{2D}^2 \approx 0.2$~\cite{BernardKrauth11}. Finally, the composite nature of the bath particles becomes relevant when the dimer size becomes comparable to the distance between dimers. 
For comparison, we also include a mean-field (MF) prediction (grey line), which estimates the energy shift based only on the reduced spatial overlap between $n_z=2$ and $n_z=0$ harmonic oscillators~\cite{supmat}. 
This trivial MF approach neglects many-body correlations intrinsic to polaron formation and breaks down in the intermediate to strongly interacting regime.

\begin{figure}[tb!]
\includegraphics[scale=1]{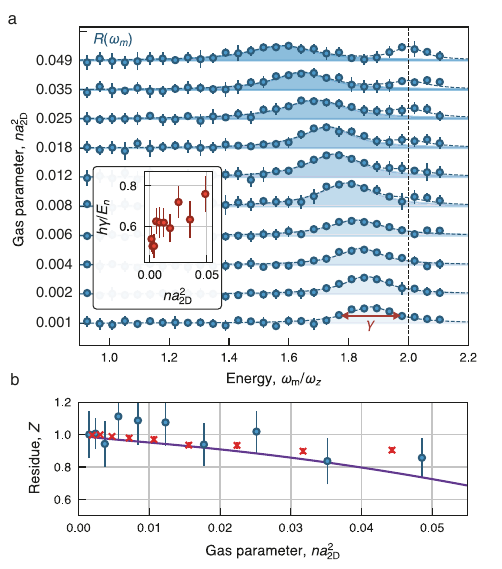}
\caption{\textbf{Spectral function at different interaction strengths.} a) Ridge plot showing the response for different $na_\textrm{2D}^2$, directly connected to the spectral function $A_2(\omega_\mathrm{m}+\mu_\mathrm{B})$. From Lorentzian fits (blue dashed lines), the full width at half maximum $\gamma$ is extracted (inset), as well as the integral, shown as blue shaded areas, which determines the quasiparticle residue $Z$. As impurity-bath interactions increase, the spectral response shifts and broadens, reflecting strong correlations with the medium. b) Quasiparticle residue $Z$, quantifying the overlap between the dressed polaron and a non-interacting particle. For increasing $na_\textrm{2D}^2$, the extracted $Z$ (blue circles) steadily decreases, signaling enhanced dressing by the bath, which is supported by QMC (red crosses) and T-matrix (purple line) calculations.}
\label{Fig:Fig3.0}
\end{figure}

Having established the energy shift $\Delta$ as a proxy for the polaron energy, we now turn to two additional quasiparticle properties encoded in the spectral response: the lifetime and the quasiparticle residue, directly related to the spectral width and weight, respectively.
As shown in Fig.~\ref{Fig:Fig3.0}a, the spectral response shows a well-defined resonance visible throughout the strongly interacting regime, indicating the remarkable stability of the repulsive polaron branch in our system.

The inset in Fig.~\ref{Fig:Fig3.0}a shows how the full width at half maximum (FWHM) $\gamma$ extracted from a Lorentzian fit remains narrow across the explored region, normalized by the characteristic energy scale $E_n=\hbar^2k_n^2/(2m)$.
This is in stark contrast to previous observations where only a broad continuum was observed in the strongly interacting regime \cite{Hu2016, Jorgensen2016,etrych2024}.
The enhanced stability arises from a fundamental distinction in how the repulsive branch is engineered. 
Unlike previous experiments, where the repulsion stems from the presence of an underlying molecular state~\cite{Scazza2020}, our scheme uses the residual repulsion between stable bosonic dimers~\cite{Petrov2004}. 
Therefore, the bound state responsible for the instability of the repulsive branch does not exist. 
This key feature not only leads to a robust, long-lived repulsive branch but also enables direct comparison with variational and quantum Monte Carlo descriptions, which typically neglect bound states~\cite{Ardila2015,Schmidt2022-rf}. 

The second quantity encoded in the spectral function is the quasiparticle residue $Z$, which quantifies how much of the original undressed particle remains after interaction with the bath: $Z=1$ for a free particle, while $Z<1$ indicates many-body dressing~\cite{Scazza2022}.
In the spectral function, $Z$ corresponds to the integrated weight of the resonance (shaded area in Fig.~\ref{Fig:Fig3.0}a), which we extract from the Lorentzian fits.
The resulting values of $Z$, normalized to the weakest interaction strength ($na_\textrm{2D}^2 \sim 0.001$), are displayed in Fig.~\ref{Fig:Fig3.0}b, showing excellent agreement with our predictions. The gradual decrease for larger interactions signals the many-body dressing by the bath, while upholding the validity of the quasiparticle picture ($Z>0$).

\begin{figure*}[htp!]
\centering
\includegraphics[width=\textwidth]{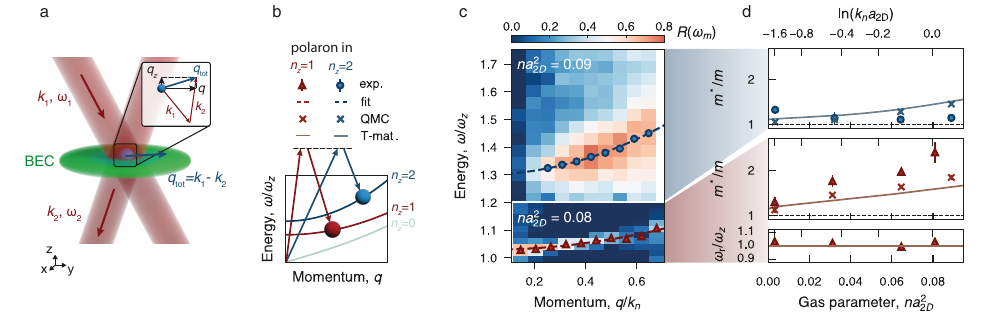}
\caption{\textbf{Effective mass of the $n_z=1$ and $n_z=2$ polaron.} a) Creation of a polaron via Bragg spectroscopy: By overlapping two beams onto the atoms, a selected energy of $\omega_\textrm{m}=\omega_1-\omega_2$ and momentum $\mathbf{q}_\textrm{tot}=\mathbf{k}_1-\mathbf{k}_2$ can be transferred. Slightly tilting the beams allows for excitations into the $n_z=1$ and $n_z=2$ impurity states due to a small transverse momentum $q_z\ll q_\textrm{tot}$ imparted in addition to an in-plane momentum $q\approx q_\textrm{tot}$ (inset). b) Bragg spectrum of a bosonic superfluid, displaying the phonon ($n_z=0$) and the impurity modes ($n_z=1$ and $n_z=2$). The inverse of the curvature of the impurity dispersion directly gives the effective mass $m^*$. c) Bragg spectrum taken in the strongly interacting regime. $E_\textrm{P}(q)$ extracted from Lorentzian fits to the momentum slices are displayed for the $n_z=1$ and $n_z=2$ mode, including the quadratic fits. d) Effective masses extracted from the quadratic fits to the Bragg spectra, including a comparison with QMC and T-matrix theory. The bottom panel shows the zero-momentum energies $\omega_\textrm{r}=E_\textrm{P}(q=0)$ for $n_z=1$ extrapolated from the fit. Remarkably, the energy shift from the bare transition remains zero for all gas parameters $na_{2D}^2$ due to symmetry.}
\label{Fig:Fig4.0}
\end{figure*}

In addition to the energy shift $\Delta$, spectral width $\gamma$, and quasiparticle residue $Z$, we measure the effective polaron mass $m^*$. This quantity characterizes the mobility of the impurity and follows the dispersion relation $E_\textrm{P}(q) = E_\textrm{P}(0)+\hbar^2 q^2 / (2m^*)$.
We measure $E_\textrm{P}(q)$ using Bragg spectroscopy, where a two-photon process transfers a precise energy $\hbar\omega_\textrm{m}$ and in-plane momentum $\hbar q$ to the impurity (see Fig.~\ref{Fig:Fig4.0}a). 
As described in our previous experiments~\cite{Sobirey2021-xy, Biss2022-ra}, the interference of two far-detuned laser beams creates a running-wave optical lattice with tunable spacing.
It is important to note that coupling into the $n_z$ states requires a small out-of-plane momentum component $q_z$ ~\cite{supmat}, which we introduce by slightly tilting the Bragg beams ($q_\textrm{tot}$) relative to the atomic plane ($q_z\approx0.05q_{\textrm{tot}}$), see inset in Fig.~\ref{Fig:Fig4.0}a.
As shown in Fig.~\ref{Fig:Fig4.0}b, this off-axis Bragg excitation allows transitions to $n_z=1$ and $n_z=2$, as the matrix elements between harmonic oscillator states are not constrained by parity~\cite{Wineland1979}. 
Using the detection protocol introduced for $\Delta$ in Fig.~\ref{Fig:figure2}, we measure the momentum-resolved resonance frequency~\cite{supmat}.

A typical Bragg spectrum in the strongly correlated regime is shown in Fig.~\ref{Fig:Fig4.0}c. 
This spectrum combines two datasets for the $n_z=1$ (bottom) and $n_z=2$ (top) impurity.
For each momentum slice, $E_\textrm{P}(q)$ is determined individually, allowing us to fit a quadratic dispersion \footnote{For extraction of the curvature, we do not include the value at $q=0$ from trap modulation spectroscopy, since there is a small systematic energy shift between both measurements.}, which gives the effective mass.
For comparison, in our theory models, we extract $m^*$ as follows:
In QMC calculations, the effective mass is obtained from the imaginary-time diffusion coefficient assuming distinguishable impurities. Independently, using T-matrix theory, we reformulate a momentum-resolved spectral function $A_{n_z}(q,\omega)$ to extract the effective mass~\cite{supmat}.
The resulting effective masses for the $n_z=1$ and $n_z=2$ polaron branches are shown in Fig.~\ref{Fig:Fig4.0}d for a wide range of $na_\textrm{2D}^2$.
In the weakly interacting regime, the measured effective masses for both $n_z=1$ and $n_z=2$ are close to $2m_{{\rm Li}}$, consistent with the picture of nearly free dimers moving through the bath. 
However, at strong coupling, the two branches noticeably differ.
For the $n_z=1$ impurity, we observe a pronounced increase in the effective mass, exceeding twice the bare dimer mass for our strongest bath-impurity coupling, which is much larger than the effective masses observed previously for Fermi polarons~\cite{Nascimbene2009,Koschorreck2012,Scazza2017,Yan2019a}.
Both T-matrix and QMC predictions qualitatively capture this increase, confirming this feature as a clear signature of many-body dressing. 
By contrast, we measure a nearly constant effective mass for the $n_z=2$ polaron.
Although $m^*$ for $n_z=2$ is expected to be smaller than for $n_z=1$ due to its reduced overlap with the bath~\cite{supmat}, the absence of any noticeable increase with stronger bath-impurity coupling is unexpected and requires further investigation.

While the effective mass reveals strong interaction effects for the $n_z=1$ impurity, the energy shift shows a remarkably different behaviour. 
The bottom panel of Fig.~\ref{Fig:Fig4.0}d shows the extrapolated zero-momentum energies $E_\textrm{P}(0)=\Delta_{n_z=1}+\hbar\omega_z$ from Bragg spectroscopy, extracted from the quadratic fit.
Surprisingly, there is no deviation from the bare transition across the full interaction range.
This striking result stems from the parity symmetry of the trapping potential: since there are no odd-parity pairwise interactions, the $n_z=1$ excitation can only go into the center-of-mass motion of the gas, without changing its properties. Within the impurity picture, this appears as a cancellation between the factor of 2 due to impurity-bath exchange (which is incorporated into our T-matrix approach) and the factor of 1/2 arising from the reduced spatial overlap between $n_z=1$ and $n_z=0$ wavefunctions. The resulting vanishing interaction energy shift \cite{supmat} thus highlights a subtle manifestation of quantum indistinguishability.

In conclusion, we have engineered a stable Bose polaron in 2D and measured the key properties from the weakly to the strongly interacting regime.
Those include the polaron energy and the quasiparticle residue, finding excellent agreement with QMC predictions and T-matrix calculations.
Furthermore, we determined the effective mass via Bragg spectroscopy, showing a significant increase in the effective polaron mass and absence of the energy shift $\Delta$ associated with the $n_z =1$ polaron, due to bosonic statistics. 
Our work demonstrates how the internal states of the confining potential can be exploited as pseudospins, enabling momentum-resolved spectroscopy without relying on RF or Raman transitions. 
We term this long-lived quasiparticle the \textit{synthetic-spin polaron}.
Next steps include extending this approach to finite temperatures, which will enable the exploration of how thermal fluctuations affect polaron stability and mobility, thereby deepening our understanding of quasiparticles in low-dimensional Bose systems.
While this work focuses on the bosonic regime, our polaron mapping, based on internal states, can be extended to imbalanced Fermi gases \cite{GUBBELS2013255}, enabling exploration of the crossover between Bose and Fermi polarons within the same experimental platform.

\textbf{Acknowledgements.}
We gratefully acknowledge fruitful discussions with L.\ Ardilla, F.\ Scazza, J.\ Arlt, G.\ Bruun, P.\ Massignan, C.\ Eigen. 
JL and MMP are supported through Australian Research Council Discovery Project DP240100569 and Future Fellowship FT200100619, respectively.
This work is supported by the Deutsche Forschungsgemeinschaft (DFG, German Research Foundation) in the framework of SFB-925—project 170620586—and the Cluster of Excellence “CUI: Advanced Imaging of Matter”—EXC 2056—project ID 390715994 and co-financed by ERDF of the European Union and by “Fonds of the Hamburg Ministry of Science, Research, Equalities and Districts (BWFGB)”.
J. B. and G. A. acknowledge financial support  from Ministerio de Ciencia e
Innovaci\'on MCIN/AEI/10.13039/501100011033
(Spain) under Grant No. PID2023-147469NB-C21.

\textbf{Author Contributions.} 
R.\,H. and C.\,R.\,C.\ performed the experiments and analysed the data.
J.\,L. and M.\,P.\ developed the T-matrix formalism. G.\,A. and J.\,B.\ performed the QMC simulations.
C.\,R.\ and H.\,M.\ supervised the work. 
All authors contributed to the interpretation of the results and to the preparation of the manuscript.

\textbf{Data availability.}
The datasets supporting this study are available from the corresponding authors upon request.

\textbf{Competing interests.} The authors declare no competing interests.

\textbf{Correspondence and requests for materials} should be addressed to C. R. Cabrera or H. Moritz.\\

\vspace{30pt}

\bibliography{RepulsivePolaron}

\setcounter{figure}{0}
\renewcommand{\thefigure}{S\arabic{figure}}
\renewcommand{\theequation}{S\arabic{equation}}
\renewcommand{\k}{\mathbf{k}}
\newcommand{\q}{\mathbf{q}}
\newcommand{\0}{\mathbf{0}}
\newcommand{\p}{\mathbf{p}}
\newcommand{\nn}{\nonumber}
\newcommand{\ek}{\epsilon_{\k}}
\newcommand{\z}{\mathbf{z}}

\clearpage
\pagebreak

\begin{center}
\textbf{\large Supplementary Material: Realization of repulsive polarons in the strongly correlated regime}
\end{center}

\section{\label{sec:prep} Reaching quantum degeneracy}
We prepare the 2D superfluid by evaporating a balanced two-component Fermi mixture of $^6$Li in the lowest two hyperfine states $\ket{F=1/2, m_F=\pm1/2}$ close to the Feshbach resonance at $\SI{832}{G}$. 
This is achieved by using an attractive, oblate dipole trap (squeeze trap) at a wavelength of $\lambda=\SI{1064}{nm}$. A magnetic field gradient deterministically spills the gas to a desired number of atoms, fluctuating less than $\SI{5}{\percent}$ in between experimental cycles. 
The atoms are then transferred to the repulsive box potential used for the main experiment: First, the magnetic field is ramped to the low-interacting bosonic regime at $\SI{750}{G}$, which shrinks the cloud for an efficient transfer into the box potential. 
Subsequently, the repulsive dipole traps for the 2D homogeneous box are switched on, consisting of the ring and accordion, both at $\lambda=\SI{532}{nm}$. The ring is projected through the microscope and radially confines the atoms within a diameter of $\sim\SI{90}{\micro m}$. For the 2D confinement, we employ a vertical accordion lattice, which is created by overlapping two beams from the side at an angle which can be continuously tuned between $\SI{1.5}{\degree}$ and $\SI{13}{\degree}$, leading to trapping frequencies ranging from $\omega_z=2\pi\cdot\SI{0.8}{kHz}$ up to $2\pi\cdot\SI{11.5}{kHz}$, respectively. 
After loading atoms into the open accordion, it is compressed, leaving a superfluid in the 2D regime. To achieve our final temperatures, a last evaporation is performed by lowering the ring power. 
For efficient evaporation, we utilize the anticonfinement of the accordion trap, which is typically canceled out by the coils that create the magnetic field for tuning the interactions. A different set of coils in Helmholtz configuration generates a homogeneous magnetic field at identical field strengths for evaporation ($\SI{832}{G}$), leaving the anticonfinement of the accordeon in place. 
Therefore, during evaporation, the hot atoms outside the ring are actively pushed away, leaving a clean superfluid at temperatures $\widetilde{T}/T_F\leq0.02$, measured after a ramp to low interactions at $na_\textrm{2D}^2=\num{6e-4}$ ($\ln(k_na_\textrm{2D})=-2.5$). All measurements in the main text are performed at a density of $n\approx\SI{0.9}{\micro m^{-2}}$ such that we remain in the 2D regime with $E_\textrm{F}<\hbar\omega_z$, where $E_\textrm{F}$ is the Fermi energy, which is an upper bound for the chemical potential $\mu_\textrm{B}$.

\section{Extraction of the Response}
For both spectroscopy measurements —trap modulation and Bragg spectroscopy —we extract the response using the following method.
When driving a resonance, the superfluid heats up, resulting in a reduced condensate density, which manifests as a reduction of the low-momentum peak in the momentum distribution.
We can access this distribution via a $T/4$ time-of-flight (TOF) measurement:
After a short rethermalization time of $\SI{20}{ms}$, we ramp the superfluid to low interactions, at a magnetic field of $B=\SI{750}{G}$, where $na_\textrm{2D}^2=\num{6e-4}$ or $\ln(k_Fa_\textrm{2D})=-2.5$.
Here, we switch off the trapping potentials and exploit the residual confinement from the magnetic field:
After a quarter of a period $T/4=\SI{9}{ms}$, the momentum space maps to real space and can be imaged via standard absorption imaging~\cite{Hueck2018-cs}.
Comparing the peak densities of the low momentum peaks of the unperturbed system $C(0)$ with the driven system $C(\omega_\textrm{m})$ then allows us to extract the response~\cite{Sobirey2021-xy}:
\begin{equation}
    R(\omega_\textrm{m})=1-\frac{C(\omega_\textrm{m})}{C(0)}.
\end{equation}

\begin{figure}[tb!]
\includegraphics[scale=1]{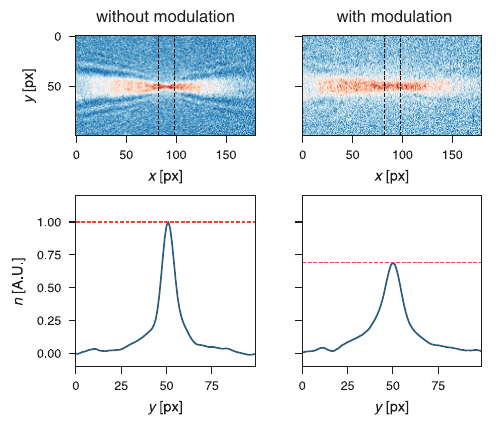}
\caption{\textbf{Extraction of the response.} Density images after a $\SI{9}{ms}$ T/4-TOF expansion for a system before (left) and after (right) resonant modulation. The imaging is slightly tilted to avoid fringes caused by atoms that have expanded out of focus from entering the central region. The smoothed vertical slices from the sum between the black dashed lines is shown in the lower panels. A clear condensate peak is visible for both slices, but with different maximum values (red dashed lines).}
\label{Fig:ST4Response}
\end{figure}

Since the superfluid is strongly confined along the axis of the imaging, the cloud rapidly expands outside the depth of field of the microscope, distorting the momentum distribution.
We resolve this issue by slightly tilting the imaging beam such that the $z$-axis is partially mapped to the $x$-axis of the density images, resulting in the elongated shape shown in the top panels of Fig.~\ref{Fig:ST4Response}. 
The center corresponds to the atoms in focus, showing the actual momentum distribution, while the atoms out of focus appear on the sides of the beetle. 
We extract the peak densities $C(\omega_\textrm{m})$ by taking a horizontal sum of the densities in between the dashed lines ($\SI{16}{px}$ wide) and smoothing the distribution, shown at the bottom of Fig.~\ref{Fig:ST4Response}.
$C(\omega_\textrm{m})$ corresponds to the maximum value (red dashed line) and is averaged over 10 images for each data point.

To extract the energy shift $\Delta$ (main text Fig. 2b), the full width at half maximum $\gamma$ (main text Fig. 3a) and the quasi-particle residue $Z$ (main text Fig. 3b), we perform a lorentzian fit to $R(\omega_\textrm{m})$ with
\begin{equation}
    f(\omega) = \frac{A}{1 + \left( \frac{\omega - \omega_\mathrm{r}}{\gamma} \right)^2}.
\end{equation}
These fits are shown in  Fig. 3a in the main text and directly give $\gamma$ and the resonance frequency $\omega_\mathrm{r}$ used to calculate $\Delta$. The integral for $Z$ is well known and yields
\begin{equation}
  \int_{-\infty}^{\infty} \frac{A}{1 + \left( \frac{\omega - \omega_\mathrm{r}}{\gamma} \right)^2} \, d\omega = A \cdot \pi \cdot \gamma.
\end{equation}
We normalize all integrals with the value at the lowest interaction, since $Z=1$ for $na_\textrm{2D}^2=0$.

The temperature is extracted from a bimodal fit to the same momentum distribution used for the response measurement, but without smoothing and performing the horizontal sum over an area roughly half as wide ($\SI{6}{px}$). 
The bimodal fit combines a parabolic term for the condensate and a Boltzmann distribution, which extracts the temperature from the thermal wings.
To achieve a stable fit, we average over 50 images.
For all measurements performed, the temperatures are at $\widetilde{T}/T_F\leq0.02$.

\section{Quasi-two-dimensional interaction parameters}
The starting point of our experiment is a Bose-Einstein condensate (BEC) of bosonic dimers composed of fermionic atoms in the two lowest hyperfine states of $^6$Li. We work in the universal regime where, in the absence of any transverse confinement, the binding energy of dimers is directly linked to their 3D $s$-wave scattering length $a_s$ via $\varepsilon_B=\frac{\hbar^2}{m_fa_s^2}$ ($m_f=m/2$ is the fermion mass). Furthermore, the dimer-dimer scattering length $a_\mathrm{3D}$ is~\cite{Petrov2004}
\begin{align}
    a_\mathrm{3D}=0.6a_s.
\end{align}
Note that, even though the dimers are extended objects of size $\sim a_\mathrm{3D}$, the dimer-dimer scattering amplitude only has a small effective range $r_\mathrm{3D}=0.13a_s$~\cite{Deltuva2017}. Thus, in a quasi-2D geometry with transverse confinement frequency $\omega_z$ it is a reasonable assumption to treat the dimers as point-like objects as long as $\varepsilon_B\gtrsim \hbar\omega_z$.

To arrive at the effective scattering properties in the confined geometry, we use the formalism of Ref.~\cite{Petrov2001-kn} (see also \cite{Bloch2008-mq,Levinsen2Dreview}). Here, we have an effective quasi-2D dimer-dimer scattering $T$ matrix at collision energy $E$: 
\begin{align}\label{eq:q2Dtmat}
    \mathcal{T}(E)=\frac{\sqrt{8\pi}\hbar^2}{m}\frac1{\frac{l_z}{a_\mathrm{3D}}-\mathcal{F}(-E/\hbar\omega_z)},
\end{align}
where $l_z=\sqrt{\hbar/m\omega_z}$ is the harmonic oscillator length and~\cite{Bloch2008-mq}
\begin{align} {\cal F}(x)=\int_0^\infty
  \frac{du}{\sqrt{4\pi u^3}}\left[1-\frac{e^{-xu}}{\sqrt{[1-
        e^{-2u}]/2u}}\right].
\label{eq:Fcali}
\end{align}

At low energy compared with the transverse confinement, the function $\mathcal{F}$ takes the form $\mathcal{F}(x)\simeq \ln(\pi x/B)/\sqrt{2\pi}$, with $B=0.905$~\cite{Petrov2001-kn,Bloch2008-mq}. Comparing with the universal low-energy form of the 2D scattering amplitude,
\begin{align}
    \mathcal{T}(E)=\frac{4\pi\hbar^2}m \frac1{\ln(\hbar^2/ma_\mathrm{2D}^2E)+i\pi},
\end{align}
then allows one to identify the effective 2D dimer-dimer scattering length
\begin{align}
    a_\mathrm{2D}=l_z\sqrt{\frac \pi B}e^{-\sqrt{\frac \pi2}l_z/a_\mathrm{3D}}.
\end{align}

The interactions between a dimer in the $n_z=0$ BEC and an impurity dimer in either the $n_z=1$ or the $n_z=2$ level is distinct from those between $n_z=0$ dimers due to the different spatial overlaps of the transverse wave functions. To see this, we first consider the simplest regime where $a_\mathrm{3D}\ll l_z$. Since in this case the dimer-dimer interactions are of much shorter range than the transverse harmonic confinement, the effective interactions are simply reduced by a factor 
\begin{align}
    \frac{\int dz\,|\varphi_{n_z}(z)|^2|\varphi_{0}(z)|^2}{\int dz\,|\varphi_{0}(z)|^4}=\left\{\begin{array}{cc} \frac12 & n_z=1 \\ & \\ \frac38 & n_z =2 \end{array}\right..
\end{align}
Beyond the regime of weak interactions, our two theories treat the interactions between impurities and medium particles slightly differently, including in terms of how they can scatter into other levels of the transverse confinement. See the discussions in Sections~\ref{sec:Tmatrix} and \ref{sec:DMC} below. 

\section{Polaron energies}
The energy shift measured via trap modulation spectroscopy is given by the polaron energy $\mu_\textrm{P}$ and the chemical potential of the bath $\mu_\textrm{B}$, such that $\Delta=\mu_\textrm{P}-\mu_\textrm{B}$, as explained in the main text. 
This protocol removes a dimer from the bath and places it in the $n_z=2$ state.
To extract $\mu_\textrm{P}$, we have to make assumptions about the bath chemical potential. 
Here we compare the following predictions:
\begin{enumerate}
    \item The chemical potential from our QMC calculations $\mu_\textrm{B,QMC}$.
    \item The mean-field (MF) prediction of 2D hard-core bosons from \cite{Schick71} corresponding to
\begin{eqnarray}
\mu_\textrm{B,MF} = \frac{4\pi\hbar^2n/m}{|\ln na_{2D}^2|}.
\label{Eq:mu:MF}
\end{eqnarray}
\item A beyond-mean-field correction to $\mu_\textrm{B,MF}$ from \cite{Astrakharchik2009}, resulting in
\begin{equation} \label{Eq:EoS:2D:BMF}
\mu_\textrm{B,BMF}  = \frac{4\pi\hbar^2n/m}{
     |\ln na_{2D}^2|
+ \ln|\ln na_{2D}^2| + C_1
+\frac{\ln|\ln na_{2D}^2| + C_2}{|\ln na_{2D}^2|}
}, 
\end{equation}
with coefficients
\begin{align*}
C_1&=-\ln\pi-2\gamma-1=-3.3 \\ 
C_2&=-\ln\pi-2\gamma+2=-0.3 .
\end{align*}
Here, $\gamma\simeq0.577$ is Euler's constant.
\end{enumerate}

\begin{figure}[tb!]
\includegraphics[scale=1]{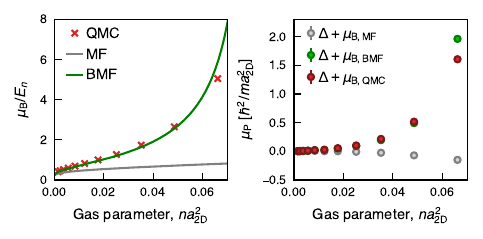}
\caption{\textbf{Comparison with various chemical potentials.} a) Comparison of different predictions for the chemical potential $\mu_\textrm{B}$ of the bath (see text). b) Polaron energy $\mu_\textrm{P}=\Delta+\mu_\textrm{B}$ assuming different $\mu_\textrm{B}$. $\Delta$ is defined as in the main text. For $na_\textrm{2D}^2<0.06$ the BMF and QMC predictions are expected to yield good results. Beyond this regime, the polaron energy is likely inaccurate. In contrast, $\mu_\textrm{B,MF}$ underestimates the chemical potential, clearly visible from the negative predicted polaron energy $\mu_\textrm{P,MF}$.}
\label{Fig:ChemicalPotentials}
\end{figure}

A comparison of the different chemical potentials is shown in Fig.~\ref{Fig:ChemicalPotentials}a, and resulting polaron energies $\mu_\textrm{P}$ in b when adding $\mu_\textrm{B}$ to the measured energy shift $\Delta$ from the experiment. 
Since our QMC theory agrees well with the results for $na_\textrm{2D}^2<0.06$ (see Fig. 2b in the main text), we expect the predicted chemical potentials to be close to the actual values for this regime. For larger interactions, the details of the scattering potential become relevant and the predictions for $\mu_\textrm{B, QMC}$ unreliable. The included BMF theory replicates these results.
For completeness, we also included the 2D hard-core MF prediction $\mu_\textrm{B,MF}$, used in the main text, which clearly underestimates the chemical potential, leading to the prediction of negative polaron energies as a consequence.
We therefore expect $\mu_\textrm{B, QMC}$ to be the most reliable prediction.

\section{Relationship between trap modulation spectroscopy and polaron spectral function}
In order to promote dimers to excited transverse levels, we modulate the transverse potential according to
\begin{align} 
    V_\perp(z) & =  \frac{1}2 m \omega_z^2 z^2 (1+ \alpha \sin(\omega_\mathrm{m} t) )^2 \nonumber \\ & \simeq \frac{1}2 m \omega_z^2 z^2 (1 + 2 \alpha \sin(\omega_\mathrm{m} t) ) \, ,
\end{align}
where the amplitude $\alpha$ of the modulation is small. This yields a time-dependent perturbation on the gas of dimers, which can be written as
\begin{equation}
    \hat{V}_\Omega(t) = 2\Omega \sin (\omega_\mathrm{m} t) \int d^3r\, z^2 \, \underbrace{\hat{\psi}^\dag (\mathbf{r}) \hat{\psi} (\mathbf{r})}_{\hat{n}(\mathbf{r})} \, ,   
\end{equation}
where $\Omega \equiv \frac{1}{2} m \omega_z^2 \alpha$ and $\hat{\psi}^\dag (\mathbf{r})$ creates a bosonic dimer at position $\mathbf{r}$. 

Now we can define
\begin{subequations} \label{eq:transform}
\begin{align}
    \hat{c}_{\k n} & = \int d^3r \, e^{-i \k. \mathbf{r}_\perp} \varphi_n^*(z) \hat{\psi} (\mathbf{r}) \\
    \hat{\psi} (\mathbf{r}) & = \sum_{\k,n} e^{i \k. \mathbf{r}_\perp} \varphi_n(z) \hat{c}_{\k n} \, ,
\end{align}
\end{subequations}
where $\varphi_n(z)$ is the harmonic oscillator wave function and $\mathbf{r}_\perp = (x,y)$. 
Thus, we have
\begin{equation}
    \hat{V}_\Omega(t) = 2\Omega \sin (\omega_\mathrm{m} t)  
    \sum_{\substack{\k, n \\ m \geq n}} W_{nm} \left(\hat{c}^\dag_{\k m} \hat{c}_{\k n} + h.c.\right) \, ,   
\end{equation}
with the coefficients
\begin{align} 
    W_{nm} & \equiv \int^\infty_{-\infty} dz \, \varphi_n^*(z) z^2 \varphi_{m}(z) \nonumber \\
    &= 
    \begin{cases}
        \frac{2n+1}2 l_z^2\, , & m=n \\
        \frac{\sqrt{(n+1)(n+2)}}{2} l_z^2 \, , & m=n+2 \\
        0 & \textrm{otherwise}
    \end{cases}.
\end{align}
These coefficients determine the amplitude for transitions from harmonic oscillator level $m$ to $n$. Here, and in the following, we set the system area and $\hbar$ to 1.

We now wish to determine the heating rate due to the trap modulation, i.e., $d \expval*{\hat{H}}/dt$, where $\hat{H}$ is the unperturbed Hamiltonian, which in the Schr\"odinger picture is given by
\begin{equation} \label{eq:ham}
    \hat{H} = \sum_{\k n} \xi_{\k n} \hat{c}^\dag_{\k n} \hat{c}_{\k n} + \hat{U}_\mathrm{int} \, , 
\end{equation}
where $\xi_{\k n} = \k^2/2m + n \omega_z$, and $\hat{U}_\mathrm{int}$ describes the interactions. Note that we measure energies with respect to the zero-point energy in the trap.
Thus, we have
\begin{align}
    i \frac{d \expval*{\hat{H}}}{dt} = \bra{\psi}U^\dag(t) [ \hat{H}, \hat{V}_\Omega(t)] U(t) \ket{\psi},
\end{align}
where $U(t)$ is the time evolution operator with respect to the full Hamiltonian $\hat{H} + \hat{V}_\Omega$, and $\ket{\psi}$ is the initial state. Now $[\hat{U}_\mathrm{int}, \hat{V}_\Omega] = 0$ since the interaction term is also a function of the density operator $\hat{n}(\mathbf{r})$. Thus, we just need to compute the commutator for the kinetic term, which yields
\begin{widetext}
\begin{align}
    [ \hat{H}, \hat{V}_\Omega(t)] = 4\Omega \sin (\omega_\mathrm{m} t)  
    \omega_z \sum_{\k, n} W_{n,n+2}  \left(\hat{c}^\dag_{\k n+2} \hat{c}_{\k n} - h.c.\right).
\end{align}

Finally, since $\hat{V}_\Omega$ is a perturbation, we expand the time evolution operator:
\begin{align} \label{eq:Uexpand}
    U(t) = U_0(t) - i U_0(t) \int_{-\infty}^t dt'\, U_0(t') 
    \hat{V}_\Omega(t') U_0(t') ,
\end{align}
where $U_0(t)= e^{-i\hat{H}t}$. After some algebra, we then obtain the heating rate up to order $\Omega^2$: 
\begin{equation}
    \frac{dE}{dt} = 8 \Omega^2 \omega_z \sin(\omega_\mathrm{m}t) \int^t_{-\infty} dt' 
    \sin(\omega_\mathrm{m} t') \sum_{\substack{\k, \k', n,n'\\ m\geq n'}} W_{n,n+2} W_{n'm} \left< \left[\left(\hat{c}^\dag_{\k' m}(t') \hat{c}_{\k' n'}(t') + h.c.\right) ,\left(\hat{c}^\dag_{\k n+2}(t) \hat{c}_{\k n}(t) - h.c.\right) \right]\right>.
\end{equation}

To make further progress, we assume that the initial state features a condensate in the $n_z=0$ level and we only consider the heating of this condensate (which is the case in the experiment). 
This allows us to make the replacement $\hat{c}_{\mathbf{0},0}(t) \simeq \sqrt{n_0} e^{-i\mu_\mathrm{B} t}$ with $n_0$ being the condensate density, which then gives
\begin{align} 
    \frac{dE}{dt} \simeq 16 \Omega^2 n_0 \omega_z \sin(\omega_\mathrm{m} t) \int^t_{-\infty} dt' 
    \sin(\omega_\mathrm{m} t') \left(2 W_{02} W_{00} \textrm{Re} \left[e^{i\mu_\mathrm{B}(t-t')}
    \expval{\hat{c}_2(t) \hat{c}_0^\dag(t')} \right]  + W_{02}^2 \textrm{Re} \left[ e^{i\mu_\mathrm{B}(t-t')}\expval{\hat{c}_2(t) \hat{c}_2^\dag(t')} \right]\right),
\end{align}
where we have used the shorthand notation $\hat{c}_{n}(t) \equiv\hat{c}_{\mathbf{0},n}(t)$.
If we neglect the off-resonant 2-0 cross term and assume the rotating wave approximation where $\omega_\mathrm{m}$ is assumed to be near $2 \omega_z$, 
we get
\begin{align} \notag
    \frac{dE}{dt} & \simeq 4 \Omega^2 n_0 \omega_z W_{02}^2 \textrm{Re} \left[ \int^t_{-\infty} dt' \, 
    e^{i(\omega_\mathrm{m} +\mu_\mathrm{B})(t-t')} \expval{\hat{c}_2(t) \hat{c}_2^\dag(t')} \right] \\
    & = 4 \Omega^2 n_0 \omega_z W_{02}^2 \textrm{Im} \underbrace{\left[ \int^{\infty}_0 dt \,
    e^{i(\omega_\mathrm{m} +\mu_\mathrm{B})t} \, i \expval{\hat{c}_2(t) \hat{c}_2^\dag(0)} \right]}_{-G_2(\omega_\mathrm{m} +\mu_\mathrm{B})}
    \, .
\end{align}
Here, $G_2(\omega)$ corresponds to the retarded Green's function for the $n_z=2$ impurity. 

Finally, since the spectral function of an $n_z=2$ impurity is $A_2(\omega) = - \textrm{Im} G_2(\omega)/\pi$, we see that the heating rate is simply proportional to $A_2$:
\begin{equation}
    \frac{dE}{dt} \propto n_0 \, A_2(\omega_\mathrm{m} + \mu_\mathrm{B}) \, ,
\end{equation}
where we have kept the condensate density $n_0$ as a reminder that the connection to the spectral function relies on the presence of an initial condensate. This explicitly shows how measuring the heating rate due to trap modulation is equivalent to standard protocols for measuring polaron properties, such as radiofrequency spectroscopy~\cite{Weizhe_PRA2020}.

\section{Bragg spectroscopy}
In the case of Bragg spectroscopy, the time-dependent perturbation becomes
\begin{equation}
    \hat{V}_\Omega(t) = \Omega \int d^3 r \, 2 \cos(\q. \mathbf{r}_\perp + q_z z - \omega t ) \,\hat{\psi}^\dag (\mathbf{r}) \hat{\psi} (\mathbf{r}) \, ,
\end{equation}
where $\Omega$ is proportional to the intensity of the applied field. Here, we take the Bragg beams to be tilted, such that the imparted momentum is both in and out of the plane, corresponding to $\q$ and $q_z$, respectively.
Using Eq.~\eqref{eq:transform}, we then obtain
\begin{equation}
    \hat{V}_\Omega(t) = \Omega \sum_{\k, n,m}
    \left(e^{-i\omega t} Q_{nm} \, \hat{c}^\dag_{\k+\q, m} \hat{c}_{\k n} +  e^{i\omega t} Q^*_{nm}  \, \hat{c}^\dag_{\k n} \hat{c}_{\k+\q, m} \right) \, ,
\end{equation}
where 
\begin{align} \label{eq:Qnm} \notag
    Q_{nm}  \equiv &\int^\infty_{-\infty} dz \, \varphi_n(z) e^{iq_z z} \varphi_{m}(z) \\ \notag
     = & \, \delta_{nm} \left(1 -\frac{q_z^2 l_z^2}{4} (2n+1) \right) + iq_z l_z \left(\sqrt{\frac{n+1}{2}}\delta_{n+1,m} + \sqrt{\frac{n}{2}}\delta_{n-1,m}\right) \\ 
    & - \frac{q_z^2 l_z^2}{4} \left(\sqrt{(n+1)(n+2)}\delta_{n+2,m} + \sqrt{n(n-1)}\delta_{n-2
,m}\right) + \ldots
\end{align}

Now, the observable in experiment is the rate of in-plane momentum transfer
\begin{equation}
    \frac{d \mathbf{P}}{dt} = -i \bra{\psi}U^\dag(t) [ \hat{\mathbf{P}}, \hat{V}_\Omega(t)] U(t) \ket{\psi} \, ,
\end{equation}
with the total in-plane momentum operator
\begin{equation}
    \hat{\mathbf{P}} = \sum_{\k n} \k \, \hat{c}^\dag_{\k n} \hat{c}_{\k n} \, .
\end{equation}
Using the fact that $\hat{\mathbf{P}}$ is conserved in the unperturbed system, i.e., $[\hat{\mathbf{P}},\hat{H}] = 0$, and 
\begin{equation}
    [ \hat{\mathbf{P}}, \hat{V}_\Omega(t)] = \Omega \q  
    \sum_{\k, n,m} \left(Q_{nm} e^{-i\omega t}\hat{c}^\dag_{\k+\q, m} \hat{c}_{\k n} - h.c.\right) \, ,
\end{equation}
and then expanding the time evolution operator $\hat{U}(t)$  
as in Eq.~\eqref{eq:Uexpand}, we finally obtain
\begin{align}
    \frac{d\mathbf{P}}{dt} = \Omega^2 \q
    \int^t_{-\infty} dt' \! 
    \sum_{\substack{\k, \k'\\ n,n',m,m'}} \left< \left[\left(Q_{n'm'}e^{-i\omega t'}\hat{c}^\dag_{\k'+\q, m'}(t') \hat{c}_{\k n'}(t') + h.c.\right) ,\left(Q_{nm} e^{-i\omega t}\hat{c}^\dag_{\k+\q, m}(t) \hat{c}_{\k n}(t) - h.c.\right) \right]\right> \, .
\end{align}
Performing the same approximation as before, where we take $\hat{c}_{\0,0}(t) \simeq \sqrt{n_0} e^{-i\mu_\mathrm{B} t}$, and then removing terms that do not conserve momentum, we have
\begin{align} \nn
        \frac{d\mathbf{P}}{dt} & \simeq 2 \Omega^2 n_0 \q \sum_n |Q_{n0}|^2\left\{ \textrm{Re} \left[ \int^t_{-\infty} dt'
    e^{i(\mu_{\rm B}+\omega)(t-t')} \expval{\hat{c}_{\q n}(t) \hat{c}_{\q n}^\dag(t')} \right] - \textrm{Re} \left[ \int^t_{-\infty} dt'
    e^{i(\mu_\mathrm{B}-\omega)(t-t')} \expval{\hat{c}_{-\q n}(t) \hat{c}_{-\q n}^\dag(t')} \right] \right\} \\ \nn
    & = 2 \Omega^2 n_0 \q \sum_n |Q_{n0}|^2 \left\{ \textrm{Re} \left[ \int^{\infty}_0 dt'
    e^{i(\mu_\mathrm{B}+\omega)t} \expval{\hat{c}_{\q n}(t) \hat{c}_{\q n}^\dag(0)} \right] - \textrm{Re} \left[ \int^{\infty}_0 dt'
    e^{i(\mu_\mathrm{B}-\omega)t} \expval{\hat{c}_{-\q n}(t) \hat{c}_{-\q n}^\dag(0)} \right] \right\} \\
    & = - 2 \Omega^2 n_0 \q \sum_n |Q_{n0}|^2 \left\{ \mathrm{Im} G_n(\q,\mu_\mathrm{B}+\omega) - \mathrm{Im} G_n(-\q,\mu_\mathrm{B}-\omega)\right\} \, ,
\end{align}
where $G_n(\q,\omega)$ corresponds to the retarded Green's function for an impurity with in-plane momentum $\q$ and harmonic-oscillator quantum number (or synthetic spin) $n$. Tuning the frequency to be near resonant with $n\omega_z$ and dropping off-resonant terms, we once again find that the response is proportional to the impurity spectral function
\begin{eqnarray}
    \frac{d\mathbf{P}}{dt} \propto n_0 \q |Q_{n0}|^2 A_n(\q, \mu_\mathrm{B}+\omega) \, .
\end{eqnarray}
This time, the spectrum is momentum resolved, allowing us to access the effective mass of the polaron. Note that the coupling to the $n \neq 0$ modes vanishes if the out-of-plane component $q_z \to 0$; thus, any measurement of the polaron effective mass requires the Bragg beams to be tilted. 
\end{widetext}

\subsection{Tilted Bragg in the experiment}
In the experiment, Bragg spectroscopy is implemented by focusing two parallel beams far from resonance at $\lambda=\SI{780}{nm}$ onto the atoms by using a high NA microscope.
The angle that controls the momentum transfer $q$ can be set by changing the distance between the beams before the microscope.
Additionally, each beam has an individual acousto-optic modulator (AOM) to introduce a frequency difference $\omega_\textrm{m}$ between the beams, leading to an energy transfer of $\hbar\omega_\textrm{m}$.
For excitation into the $n_z=1$ and $n_z=2$ states, an out-of-plane momentum $q_z$ is introduced by slightly misplacing the center of the beams by $\sim\SI{1}{mm}$ from the center of the microscope, such that $q_z$ has only $\sim\SI{5}{\percent}$ of the total momentum.
Using~\eqref{eq:Qnm}, we can compare the coupling within the ground state to the coupling of the $n_z=1$ and $n_z=2$ impurity states:
\begin{align} \notag
    |Q_{00}|^2 = & \, \left(1 -\frac{q_z^2 l_z^2}{4}\right)^2 \\
    |Q_{10}|^2 = & \, \frac{q_z^2 l_z^2}{2} \notag\\ 
    |Q_{20}|^2 = & \, \frac{q_z^4 l_z^4}{8}
\end{align}

\begin{widetext}
    
\section{$T$ matrix description of the polaron quasiparticle}
\label{sec:Tmatrix}

\subsection{Interactions in the quasi-2D system}
Let us start by considering  the interaction term $\hat{U}_\mathrm{int}$ in the quasi-2D Hamiltonian in Eq.~\eqref{eq:ham}, which is the standard boson-boson interaction in 3D:
\begin{equation}
    \hat{U}_\mathrm{int} = \frac{1}2\int d^3r\int d^3r'\, V(\mathbf{r} - \mathbf{r}') \, \hat{\psi}^\dag (\mathbf{r}) \hat{\psi}^\dag (\mathbf{r}') \hat{\psi} (\mathbf{r}')\hat{\psi} (\mathbf{r}) \, ,
\end{equation}
where $V(\mathbf{r})$ is the interaction potential between bosons. Using Eq.~\eqref{eq:transform} to write it in the harmonic-oscillator basis gives
\begin{equation}
    \hat{U}_\mathrm{int} = \frac{1}{2} \sum_{\k,\k',\q}\sum_{\substack{n_1, n_2 \\ n_3, n_4}} V^{n_1n_2}_{n_3n_4} \, \hat{c}^\dag_{\k, n_1} \hat{c}^\dag_{\k', n_2}  \hat{c}_{\k'+\q, n_3} \hat{c}_{\k'-\q, n_4} \, ,
\end{equation}
where
\begin{equation}
    V^{n_1n_2}_{n_3n_4} = g \sqrt{2\pi}l_z \sum_{N \nu\nu'} f_\nu 
    \braket{n_1n_2}{N\nu}  f_{\nu'}
    \braket{N\nu'}{n_3n_4} \, ,
\end{equation}
and we have assumed short-range interactions with strength $g$. Here $\braket{n_1n_2}{N\nu}$ is the (real) Clebsch-Gordan coefficient to go from the HO motion of two particles in the $n_1$ and $n_2$ levels to the centre-of-mass and relative motion quantum numbers $N$ and $\nu$, respectively~\cite{SMIRNOV1962346}. This is zero unless $n_1+n_2=N+\nu$. Also, the coefficients $f_\nu$ are only non-zero if $\nu$ is even and they correspond to the 
HO wave function of the relative motion at zero separation: $|f_\nu|^2=\frac1{\sqrt{2\pi}l_z}\frac{(\nu-1)!!}{\nu!!}$.

To understand how factors of 2 appear due to exchange, consider the  simple non-interacting state involving an $n=0$ condensate of density $n_0$ and a single boson in the $n\neq0$ level:
\begin{equation}
    \ket{\Psi} = \hat{c}^\dag_{\0, n} \ket{\textrm{BEC}_0} .
\end{equation}
This is the starting state for any polaron theory. Then we have the expectation value:
\begin{equation}
    \expval{\hat{U}_\mathrm{int}}{\Psi} = \frac{1}{2} n_0 \left[V^{0n}_{0n} + V^{n0}_{0n} + V^{0n}_{n0} + V^{n0}_{n0} \right] = 2 n_0 V^{0n}_{0n} \, ,
\end{equation}
since the ordering of indices in $V^{n_1n_2}_{n_3n_4}$ does not change the coefficient. Thus, we automatically arrive at a factor of 2 due to the exchange of bosons. Note that we can get something different from a factor of 2 if the ordering of indices does change the interaction coefficient, which would happen if the range of the potential is important. This is likely to be the case for the interaction between composite bosons.

\subsection{General formalism for spectroscopy}
The Green's function for an impurity particle (i.e., a particle that has been removed from the condensate) satisfies the Dyson equation
\begin{align}
    G_{nm}(\k,\omega)=\left[G^{(0)}_{nm}(\k,\omega)^{-1}-\Sigma_{nm}(\k,\omega)\right]^{-1}.
\end{align}
$n$ and $m$ are the incoming/outgoing harmonic oscillator indices of the transverse motion. In the absence of interactions, this index is preserved and we have
\begin{align}
    G_{nm}^{(0)}(\k,\omega)=\frac{\delta_{nm}}{\omega-n \omega_z-\ek}\equiv G_{n}^{(0)}(\k,\omega)\delta_{nm}.
\end{align}
$\ek=k^2/2m$ is the bare impurity dispersion. The question is which approximations we should apply to arrive at the self energy $\Sigma$, and thus the spectral function.

Assuming that the response is dominated by two-body correlations, the self energy due to an impurity interacting with a condensate in the ground state is (see \cite{Levinsen2012a})
\begin{align}\label{eq:Sigmacomplicated}
    \Sigma_{nm}(\k,\omega)=2n_0 T_{m0}^{n0}(\k,\omega),
\end{align}
where the factor 2 arises from exchange, as discussed above. Here, $T_{n0}^{m0}$ is the $T$ matrix for an incoming/outgoing pair of particles with HO indices $\{n,0\}$ and $\{m,0\}$.

According to Refs.~\cite{Pietila2012b,Levinsen2012a,Levinsen2Dreview}, we go into the CM frame (scattering is independent of the CM motion) via
\begin{align}\label{eq:Tmatcomplicated}
    T_{n0}^{m0}(\k,\omega)&=\sum_{N,\nu,\nu'} \sqrt{2\pi}l_z f_\nu f_{\nu'} \braket{n0}{N\nu}\mathcal{T}(\omega-N\omega_z-\ek/2)\braket{N\nu'}{m0},
\end{align}
in terms of the quasi-2D $T$ matrix defined in Eq.~\eqref{eq:q2Dtmat}. 

\subsection{Approximating the number of involved levels}
In Bragg spectroscopy, we measure the spectral functions $A_1$ and $A_2$, while in trap modulation spectroscopy we measure $A_2$. Furthermore, the frequency is always close to either $\omega_z$ or $2\omega_z$. It is therefore reasonable to consider the approximations
\begin{align}
    G_{1}(\k,\omega) \equiv G_{11}(\k,\omega)&\simeq \frac1{\omega-\ek-\omega_z-\Sigma_{11}(\k,\omega)} \\
    G_{2}(\k,\omega) \equiv G_{22}(\k,\omega)&\simeq \frac1{\omega-\ek-2\omega_z-\Sigma_{22}(\k,\omega)}.
\end{align}
We have checked that this approximation works extremely well for the results presented in this work.

\subsection{Impurity self energy}
\subsubsection{$n=2$}

For the $n=2$ impurity, we are interested in $\Sigma_{22}(\k,\omega)=2n_0 T_{20}^{20}(\k,\omega)$, in which case there are two non-vanishing Clebsch-Gordan coefficients that allow one to go from individual particles with harmonic oscillator indices 0 and 2 into the CM frame:
\begin{align}
    \braket{N=2,\nu=0}{n_1=0,n_2=2}=\braket{N=0,\nu=2}{n_1=0,n_2=2}=1/2.
\end{align}
Furthermore, we have
\begin{align}
    \sqrt{2\pi}l_z f_0^2=1,\qquad \sqrt{2\pi}l_z f_2^2=1/2.
\end{align}
Putting everything together using Eqs.~\eqref{eq:Sigmacomplicated} and \eqref{eq:Tmatcomplicated}, we find
\begin{align}
    \Sigma_{22}(\k, \omega)=2n_0\left[\frac18\frac{\sqrt{2\pi}}{m_r}\frac1{\frac{l_z}{a_s}-\mathcal{F}(-\omega/\omega_z+\ek/2\omega_z)}+\frac14\frac{\sqrt{2\pi}}{m_r}\frac1{\frac{l_z}{a_s}-\mathcal{F}(-\omega/\omega_z+2+\ek/2\omega_z)} \right].
\end{align}
The non-trivial energy dependence in $\mathcal{F}$ is due to the facts that the harmonic oscillator energy of the two scattering particles can go into either the relative or the CM motion, and that the collision energy only depends on the relative motion.

In the limit of weak interactions, $l_z/a_s\gg1$, this reduces to 
\begin{align}
    \Sigma_{22}(\k,\omega)=2n_0\frac38\frac{\sqrt{2\pi}a_s}{m_rl_z}\equiv E_\mathrm{MF,2}.
\end{align}
This is independent of the energy itself, and corresponds to the mean-field energy shift $E_\mathrm{MF,2}$. We can also arrive at this expression by taking an effective quasi-2D interaction constant
\begin{align}
    \frac{2\pi a_s}{m_r}\int dz\, |\varphi_0(z)|^2|\varphi_2(z)|^2=\frac38\frac{\sqrt{2\pi} a_s}{m_rl_z}.
\end{align}

\subsubsection{$n=1$}
For the $n=1$ impurity, we instead have
\begin{align}
    \Sigma_{11}(\k, \omega)=2n_0\frac12\frac{\sqrt{2\pi}}{m_r}\frac1{\frac{l_z}{a_s}-\mathcal{F}(-\omega/\omega_z+1+\ek/2\omega_z)},
\end{align}
where we again use Eqs.~\eqref{eq:Sigmacomplicated} and \eqref{eq:Tmatcomplicated}.
The associated mean-field energy shift is $E_\mathrm{MF,1}=n_0\frac{\sqrt{2\pi}a_s}{m_rl_z}$.

Remarkably, the self energy for the $n=1$ level appears as though the impurity-medium interactions are precisely those between two medium particles  due to a cancellation between the factor 2 from exchange and a factor 1/2 from the Clebsch-Gordan coefficient: 
\begin{align}
    \braket{N=1,\nu=0}{n_1=0,n_2=1}=1/2.    
\end{align}
Therefore,  we expect no interaction-induced energy shift at zero momentum. This is precisely what is observed in our experiment. 

\subsection{Approximating repulsive interactions}
Since the quasi-2D resummation of the interactions in Refs.~\cite{Petrov2001-kn,Bloch2008-mq} applies to attractive interactions, we need some way of making this closer to a true repulsive interaction. In particular, we note that for a repulsive interaction we always evaluate the self energy at a shifted energy, i.e., we take $\Sigma(\omega)\to \Sigma(\omega-E_\mathrm{MF})$, with $E_\mathrm{MF}$ 
the mean-field energy for either $n=1$ or $n=2$. This ensures that the repulsive branch is shifted up, as it should be for a true repulsive interaction.

We therefore find the impurity spectral functions
\begin{align}
    A_1(\k,\omega)& =-\frac1\pi \mathrm{Im}\frac1{\omega-\omega_z-\epsilon_\k-\Sigma_{11}(\k,\omega-E_\mathrm{MF,1})},\\
    A_2(\k,\omega) & =-\frac1\pi \mathrm{Im}\frac1{\omega-2\omega_z-\epsilon_\k-\Sigma_{22}(\k,\omega-E_\mathrm{MF,2})}.
\end{align}
These are then used to extract the polaron energies, effective masses, and residues shown in the main text. Note that when comparing with $\Delta$, we use $\mu_\mathrm{B}$ extracted from QMC, see below.

\end{widetext}

\section{Monte Carlo simulation details}
\label{sec:DMC}

The quantum Monte Carlo (QMC) algorithm is well-suited for computing the ground-state energy of bosonic systems. It can also be extended to evaluate the energy of specific excited states, provided their nodal structure is known. Below, we summarize the key modifications required to treat the $n=1$ and $n=2$ excitations.

We consider a ground-state guiding wave function of the form
\begin{equation}
\Psi_0({\bf r}_1,\dots,{\bf r}_N)
= \prod_{i=1}^N f_1({\bf r}_i)
\prod_{i<j} f_2(|{\bf r}_i - {\bf r}_j|),
\label{Eq:Psi0}
\end{equation}
where the one-body term $f_1({\bf r})$ accounts for the external harmonic confinement along the $z$-direction:
\begin{equation}
f_1({\bf r}) = \exp(-\alpha z^2 / l_z^2),
\label{Eq:f1}
\end{equation}
with $\alpha$ being a variational parameter whose value is optimized by minimizing the variational energy. For small values of the gas parameter, $\alpha = 1/2$ corresponds to the single-particle ground state of a harmonic oscillator, $f_1(z) = H_0(z)$. The two-body Jastrow term $f_2(r)$ is constructed by matching the short-range two-body scattering problem at short distances to long-range phononic asymptotics.

To describe an excited state, we multiply the nodeless ground-state wave function $\Psi_0$ by a nodal $N$-body term $f_N$:
\begin{equation}
\Psi({\bf r}_1,\dots,{\bf r}_N) = \Psi_0({\bf r}_1,\dots,{\bf r}_N) \, f_N({\bf r}_1,\dots,{\bf r}_N).
\label{Eq:Psi}
\end{equation}

The drift force acting on particle $i$ gets an additional $N$-body contribution,
\begin{equation}
{\bf F}_i = \frac{\nabla_i \Psi}{\Psi} = 
\frac{\nabla_i \Psi_0}{\Psi_0} + \frac{\nabla_i f_N}{f_N}
= {\bf F}^{(0)}_i + {\bf F}^{(N)}_i.
\label{Eq:Fi:n=1}
\end{equation}
Also, the contribution of particle $i$ to the kinetic energy has additional terms,
\begin{align*}
T_i & = -\frac{\hbar^2}{2m} \frac{\Delta_i \Psi}{\Psi} \\
& = -\frac{\hbar^2}{2m} \left[
\frac{\Delta_i \Psi_0}{\Psi_0}
+ 2\frac{\nabla_i \Psi_0}{\Psi_0} \cdot \frac{\nabla_i f_N}{f_N}
+ \frac{\Delta_i f_N}{f_N}
\right].
\end{align*}
Summing over all particles, the total kinetic energy reads
\begin{equation}
T = T_0 - 2\frac{\hbar^2}{2m} \sum_i {\bf F}_i^{(0)} \cdot {\bf F}_i^{(N)} 
- \frac{\hbar^2}{2m} \sum_i \frac{\Delta_i f_N}{f_N},
\label{Eq:T}
\end{equation}
where $T_0$ is the contribution to the kinetic energy calculated with respect to the ground-state wave function.

For an $n=1$ excitation, particle $i$ can occupy the first excited harmonic oscillator state, $H_1(z_i) \sim z_i \exp(-z_i^2/2)$, which corresponds to multiplying $\Psi_0$ by $z_i$. Symmetrizing over all particles leads to the following choice,
\begin{equation}
f_N({\bf r}_1,\dots,{\bf r}_N) = \sum_{i=1}^N \left(z_i - \langle z\rangle\right),
\label{Eq:fN:n=1}
\end{equation}
which introduces a single node when the center-of-mass coordinate $z_{\text{CM}} = \sum_i z_i / N $ coincides with its average position, which in the considered case corresponds to the center of the trap, $\langle z \rangle = 0$.
The excitation contribution to the drift force is
\begin{equation}
{\bf F}_i^{(N)} = \frac{\nabla_i f_N}{f_N} = \frac{1}{f_N} \hat{z},
\end{equation}
which diverges when $f_N = 0$, i.e., at $z_{\text{CM}} = 0$. This divergence ensures that such configurations are not sampled during the Monte Carlo simulation.
In contrast, the second derivatives vanish, $\Delta_i f_N = 0$, and thus the last term in Eq.~\eqref{Eq:T} drops out for $n=1$ excitation.
The only contribution comes from the product of drift forces, in which the drift force~(\ref{Eq:Fi:n=1}) describing the excitation has elements only along $z$ and all of them are equal to $1/f_N$, so that for $n=1$
\begin{equation}
T = T_0 - 2 \frac{\hbar^2}{2m}\sum_i \frac{F_{i,z}^{(0)}}{f_N}
\label{Eq:T:n=1}
\end{equation}

For an $n=2$ excitation, particle $i$ can occupy the second excited harmonic oscillator state, $H_2(z_i) \sim (z_i^2-1/2)\exp(-z_i^2/2)$, which corresponds to multiplying $\Psi_0$ by $(z_i^2-1/2)$ (here we use dimensionless notation corresponding to the harmonic oscillator units). We consider the following symmetrized $N$-body term
\begin{equation}
f_N({\bf r}_1,\dots,{\bf r}_N) = \sum_{i=1}^N (z_i^2 -\langle z^2\rangle)
\label{Eq:fN:n=2}
\end{equation}
where the sum corresponds to the instantaneous average value of mean square dispersion and we subtract its average value. This wave function has two nodes.

The excitation contribution to the drift force is
\begin{equation}
{\bf F}_i^{(N)} = \frac{2 z_i}{f_N} \hat{z},
\end{equation}
and to the kinetic energy is
\begin{equation}
T = T_0 - 2\frac{\hbar^2}{m}\sum\limits_{i=1}^N \frac{F_{i,z}^{(0)} z_i}{f_N}-\frac{\hbar^2}{m}\frac{N}{f_N}
\label{Eq:T:n=2}
\end{equation}
The first odd excitation ($n=1$) imposes a node in the center-of-mass coordinate at $z_{CM} = 0$, which, on average, corresponds to the central position of the trap, $\z = \langle z_{CM} \rangle = 0$.
This location of the node is natural, as the wave function of the first excited state must be orthogonal to the ground state. The resulting density profiles for the $n=0$ and $n=1$ states are symmetric with respect to $z = 0$, while the wave function for $n=1$ is antisymmetric.
This property holds for any interaction strength and allows one to carry out QMC simulations of the $n=1$ state. In contrast, the nodal surface imposed for the $n=2$ state assumes that the ground-state density corresponds to that of a non-interacting Bose gas, which is valid in the weakly interacting regime.

To verify that the proposed many-body wave functions correctly describe the $n=1$ and $n=2$ excitations, we test them on one-dimensional ideal gases confined in a harmonic trap—namely, the ideal Bose gas with $f_1(z) = \exp(-z^2/2)$ and $f_2(z) = 1$, and the ideal Fermi gas with the same $f_1(z)$ and $f_2(z) = z$.
It can be explicitly verified that the ground-state wave function~(\ref{Eq:Psi0}) and the first two excited states~(\ref{Eq:Psi}) are exact for any number of particles, taking into account that $\langle z^2\rangle = 1/2$ for ideal bosons and $\langle z^2\rangle = N/2$ for ideal fermions. The corresponding excitation energies are $\hbar\omega$ and $2\hbar\omega$ for the $n=1$ and $n=2$ excitations, respectively. 
As we are interested in the quasi-two-dimensional regime, where the density profile is similar to the Gaussian one of the ideal Bose gas, in the calculations we assume $\langle z^2\rangle = 1/2$. 

For the excitation with a finite momentum, we consider a different symmetric many-body term that introduces a node in the $x$-component of the center-of-mass (CM) coordinate. Specifically, we multiply the nodeless wave function by
\begin{equation}
f_N({\bf r}_1,\dots,{\bf r}_N) = \cos\left( k_\ell x_{\text{CM}} \right),
\label{Eq:fN:cos}
\end{equation}
$k_\ell = \pi \ell / L$ with $\ell = 0;\pm 1; \pm 2,...$ defines the excitation momentum and $x_{\text{CM}} = \frac{1}{N} \sum_{i=1}^N x_i$ is the CM coordinate along the in-plane ($x$) direction.

The gradient with respect to particle $i$ is:
\begin{equation}
\nabla_i f_N = \frac{\partial f_N}{\partial x_i} \hat{x}
= -\frac{k_\ell}{N} \sin\left(k_\ell x_{\text{CM}} \right) \hat{x},
\end{equation}
so the drift force contribution becomes:
\begin{equation}
{\bf F}_i^{(N)} = \frac{\nabla_i f_N}{f_N}
= -\frac{k_\ell}{N} \tan\left(k_\ell x_{\text{CM}} \right) \hat{x}.
\end{equation}

The Laplacian of $f_N$ with respect to $x_i$ is:
\begin{equation}
\Delta_i f_N = \frac{\partial^2 f_N}{\partial x_i^2}
= -\left( \frac{k_\ell}{N} \right)^2 \cos\left(k_\ell x_{\text{CM}} \right),
\end{equation}
and dividing by $f_N$, we obtain:
\begin{equation}
\frac{\Delta_i f_N}{f_N} 
= -\left(\frac{k_\ell}{N} \right)^2.
\end{equation}

Summing over all particles, the total additional contribution to the kinetic energy is:
\begin{equation}
T = T_0 + \frac{\hbar^2k_\ell}{Nm}\sum_{i=1}^N F_{xi}^{(0)}\tan(k_\ell x_{CM})
+ \frac{\hbar^2}{2Nm} k_\ell^2.
\end{equation}
The contribution from only the last term corresponds to a center-of-mass excitation with an effective mass $m^\star = Nm$.
\end{document}